\documentclass[a4paper,UKenglish,cleveref,autoref]{lipics-v2021}
\title{Induced Disjoint Paths Without an Induced Minor}
\titlerunning{Induced Disjoint Paths Without an Induced Minor}

\author{Pierre Aboulker}{DIENS, \'Ecole normale sup\'erieure, CNRS, PSL University, Paris, France \and \url{https://www.di.ens.fr/~paboulker/} }{pierreaboulker@gmail.com}{}{}
\author{\'{E}douard Bonnet}{CNRS, ENS de Lyon, Université Claude Bernard Lyon 1, LIP UMR 5668, Lyon, France \and \url{http://perso.ens-lyon.fr/edouard.bonnet}}{edouard.bonnet@ens-lyon.fr}{https://orcid.org/0000-0002-1653-5822}{}
\author{Timothé Picavet}{LaBRI, Université de Bordeaux, France}{timothe.picavet@u-bordeaux.fr}{https://orcid.org/0000-0002-7129-0127}{}
\author{Nicolas Trotignon}{CNRS, ENS de Lyon, Université Claude Bernard Lyon 1, LIP UMR 5668, Lyon, France \and \url{http://perso.ens-lyon.fr/nicolas.trotignon}}{nicolas.trotignon@ens-lyon.fr}{}{}

\authorrunning{P. Aboulker, \'E. Bonnet, T. Picavet, N. Trotignon}

\Copyright{Pierre Aboulker, Édouard Bonnet, Timothé Picavet, Nicolas Trotignon}%TODO mandatory, please use full first names. LIPIcs license is "CC-BY";  http://creativecommons.org/licenses/by/3.0/

\category{}%optional, e.g. invited paper

\relatedversion{}%optional, e.g. full version hosted on arXiv, HAL, or other respository/website
\supplement{}%optional, e.g. related research data, source code, ... hosted on a repository like zenodo, figshare, GitHub, ...

\acknowledgements{We thank Théo Pierron and Jean-Florent Raymond with whom this project started. We also thank Tuukka Korhonen, Daniel Lokshtanov, and Martin Milanič for some discussions, and in particular for asking about the flow problem (as opposed to the linkage one).}%optional

\nolinenumbers %uncomment to disable line numbering

\hideLIPIcs  %uncomment to remove references to LIPIcs series (logo, DOI, ...), e.g. when preparing a pre-final version to be uploaded to arXiv or another public repository

\EventEditors{John Q. Open and Joan R. Access}
\EventNoEds{2}
\EventLongTitle{42nd Conference on Very Important Topics (CVIT 2016)}
\EventShortTitle{CVIT 2016}
\EventAcronym{CVIT}
\EventYear{2016}
\EventDate{December 24--27, 2016}
\EventLocation{Little Whinging, United Kingdom}
\EventLogo{}
\SeriesVolume{42}
\ArticleNo{23}

\usepackage[utf8]{inputenc}  

\usepackage[T1]{fontenc}
\usepackage{lmodern}

\usepackage{amsmath}  
\usepackage{amssymb}
\usepackage{amsthm}
\usepackage{bbm}
\usepackage{accents}
\usepackage{complexity}

\usepackage{booktabs}
\usepackage{paralist}
\usepackage{fixmath}

\makeatletter
\newtheorem*{rep@theorem}{\rep@title}
\newcommand{\newreptheorem}[2]{%
\newenvironment{rep#1}[1]{%
 \def\rep@title{#2 \ref{##1}}%
 \begin{rep@theorem}}%
 {\end{rep@theorem}}}
\makeatother

\newreptheorem{theorem}{Theorem}
\newreptheorem{lemma}{Lemma}
\newreptheorem{corollary}{Corollary}

\usepackage{pgfplots}
\usepackage{xspace}
\usepackage{tikz}
\usepackage{tikz-3dplot}

\usepackage[ruled,vlined,linesnumbered]{algorithm2e}

\usetikzlibrary{fit}
\usetikzlibrary{arrows}
\usetikzlibrary{patterns}
\usetikzlibrary{calc}
\usetikzlibrary{shapes}
\usetikzlibrary{positioning}
\usetikzlibrary{math}
\usetikzlibrary{shapes.symbols, shapes.geometric}
\usetikzlibrary{decorations.pathreplacing,calligraphy}
\usetikzlibrary{decorations.pathmorphing, backgrounds}
\usepackage[scr=boondox,scrscaled=1.05]{mathalfa}

\newenvironment{proofofclaim}{\noindent \textsc{Proof of the Claim:}}{\unskip\nobreak\hfill$\Diamond$\medskip}

\begin{document}

\maketitle

\begin{abstract}
  We exhibit a new obstacle to the nascent algorithmic theory for classes excluding an induced minor.
  We indeed show that on the class of string graphs---which avoids the 1-subdivision of, say, $K_5$ as an induced minor---\textsc{Induced 2-Disjoint Paths} is NP-complete.
  So, while \textsc{$k$-Disjoint Paths}, for a~fixed $k$, is polynomial-time solvable in general graphs, the absence of a~graph as an induced minor does not make its induced variant tractable, even for $k=2$.
  This answers a~question of Korhonen and Lokshtanov [SODA~'24], and complements a~polynomial-time algorithm for \textsc{Induced $k$-Disjoint Paths} in classes of bounded genus by Kobayashi and Kawarabayashi [SODA~'09].
  In addition to being string graphs, our produced hard instances are subgraphs of a~constant power of bounded-degree planar graphs, hence have bounded twin-width and bounded maximum degree.
  
  We also leverage our new result to show that there is a fixed subcubic graph $H$ such that deciding if an input graph contains $H$ as an induced subdivision is NP-complete.
  Until now, all the graphs $H$ for which such a~statement was known had a~vertex of degree at~least~4.
  This answers a~question by Chudnovsky, Seymour, and the fourth author [JCTB '13], and by Le [JGT '19].
  Finally we resolve another question of Korhonen and Lokshtanov by exhibiting a~subcubic graph $H$ without two adjacent degree-3 vertices and such that deciding if an input $n$-vertex graph contains $H$ as an induced minor is NP-complete, and unless the Exponential-Time Hypothesis fails, requires time $2^{\Omega(\sqrt n)}$.
  This complements an algorithm running in subexponential time $2^{\Tilde{O}(n^{2/3})}$ by these authors [SODA~'24] under the same technical condition.
  \end{abstract}

\section{Introduction}\label{sec:intro}

In \textsc{$k$-Disjoint Paths}, one is given a~graph $G$, together with $k$ pairs of vertices, often called \emph{terminals}, $(s_1,t_1), \ldots, (s_k,t_k)$, and has to decide if $G$ admits $k$ vertex-disjoint paths $P_1, \ldots, P_k$ such that for every $i \in \{1, \ldots, k\}$, the endpoints of $P_i$ are $s_i$ and $t_i$.
This problem is also called \textsc{$k$-Linkage} as the pairs to connect are prescribed.
In the ``flow variant'' of this problem, \textsc{Disjoint $S$--$T$ Paths}, the input consists of a~graph $G$ and two disjoint vertex subsets $S, T \subset V(G)$ with $|S|=|T|$, and the question is whether there are $|S|$ vertex-disjoint paths, each with one endpoint in $S$ and the other endpoint in $T$.
These problems are polynomial-time solvable in general graphs, respectively by the work of Robertson and Seymour~\cite{Robertson95} (for \textsc{$k$-Disjoint Paths}), and simply by Menger's theorem~\cite{Menger27} (for \textsc{Disjoint $S$--$T$ Paths}).

In this paper, we are interested in their induced variants \textsc{Induced $k$-Disjoint Paths} and \textsc{Induced Disjoint $S$--$T$ Paths}, where the paths are further requested to be mutually induced (i.e., no edge of the graph is incident to two of these paths).
These problems are NP-complete for $k=2$ and for $|S|=|T|=2$, respectively, in general graphs~\cite{Fellows89,bienstock:evenpair}.
Note that \textsc{Induced Disjoint $S$--$T$ Paths} with $|S|=|T|=k$ can be solved with $k!$ calls to \textsc{Induced $k$-Disjoint Paths}.
So when $k$ is constant, the former problem is in principle simpler than the latter.

Kawarabayashi and Kobayashi give a~linear-time algorithm for \textsc{Induced $k$-Disjoint Paths} (for any fixed $k$) in planar graphs~\cite{KawarabayashiK12}, and a~polynomial-time algorithm in classes of bounded genus~\cite{KobayashiK09}.
The latter result is lifted to the model checking of the FO+SDP logic (i.e., first-order with \emph{scattered} disjoint paths predicates, which can natively express the existence of a~constant number of induced disjoint paths between some specified pairs of terminals) in fixed-parameter tractable (FPT) time in classes of bounded genus~\cite{GolovachST23}.
It is believed that the tractability of \textsc{Induced $k$-Disjoint Paths} (and perhaps even of FO+SDP model checking) holds more generally in classes excluding a~fixed minor, and could be shown by combining the irrelevant-vertex techniques (developed for the planar and bounded-genus cases) with the graph structure theorem of Robertson and Seymour~\cite{Robertson03}. 
However, to our knowledge, this has not been proven yet (provided it indeed holds).

\textsc{Induced $k$-Disjoint Paths} is polynomial-time solvable on classes of bounded mim-width on which mim-width can be efficiently approximated~\cite{Jaffke20}, which includes for instance interval graphs and permutation graphs.
In claw-free graphs (i.e., graphs excluding $K_{1,3}$ as an induced subgraph), \textsc{Induced $k$-Disjoint Paths} is solvable in polynomial time for any fixed $k$~\cite{DBLP:journals/algorithmica/FialaKLP12}, and even in FPT time in parameter~$k$~\cite{DBLP:journals/siamdm/GolovachPL15}.
In graphs without asteroidal triple, this problem is polynomial-time solvable even if~$k$ is part of the input~\cite{GolovachPL22}.
Finally, in (theta, wheel)-free graphs, there is a~polynomial-time algorithm for~\textsc{Induced $k$-Disjoint Paths} \cite{RadovanovicTV21}, while its complexity in theta-free graphs is open.

To understand better the tractability frontier of \textsc{Induced $k$-Disjoint Paths}, Korhonen and Lokshtanov~\cite{KorhonenL23} ask if this problem is NP-hard in \mbox{$H$-induced}-minor-free graphs for some fixed $k$ and $H$.
We resolve this question already in the case when $k=2$ and $H$ is equal to the 1-subdivision of~$K_5$ (or of~$K_{3,3}$).
Indeed string graphs (i.e., intersection graphs of curves in the plane) exclude the 1-subdivision of any non-planar graph as an induced minor.  

\begin{theorem}\label{thm:hardness-I2DP}
  \textsc{Induced 2-Disjoint Paths} is NP-complete in string graphs that are subgraphs of a~constant power of bounded-degree planar graphs.
\end{theorem}

We actually show the stronger result that \textsc{Induced Disjoint $S$--$T$ Paths} with $|S|=|T|=2$ is NP-complete in this subclass of string graphs. 
Note that subgraphs of constant powers of bounded-degree planar graphs both have bounded twin-width~\cite{twin-width1,twin-width2} and bounded maximum degree.
Thus \cref{thm:hardness-I2DP} considerably limits the scope within which the existing polynomial algorithms could be extended.

The Exponential-Time Hypothesis (ETH for short), a~stronger assumption than \mbox{P $\neq$ NP} but still widely believed, asserts that there is a~real $\lambda>1$ such that $n$-variable \textsc{3-SAT} cannot be solved in time~$O(\lambda^n)$~\cite{Impagliazzo01}.
More quantitatively, \cref{thm:hardness-I2DP}, by providing a~linear reduction from a~variant of \textsc{Planar 3-SAT} (see~\cite{Lichtenstein82}), implies the following.
\begin{corollary}\label{cor:tww}
  \textsc{Induced Disjoint $S$--$T$ Paths} with $|S|=|T|=2$ is NP-complete in string graphs of bounded maximum degree and twin-width, and requires time $2^{\Omega(\sqrt n)}$ on $n$-vertex such graphs, unless the Exponential-Time Hypothesis fails.
\end{corollary}

Our result has some consequences for the detection of induced subdivisions, which we now turn our attention to.

\paragraph*{Detecting induced subdivisions and induced minors}

Let \textsc{$H$-Induced Subdivision Containment} (\textsc{$H$-ISC} for short) input a~graph~$G$ and ask whether $H$ is an induced subdivision of~$G$.
This problem has attracted some attention.
Chudnovsky and Seymour introduced the \textsc{Three-In-A-Tree} problem---whether there is an induced subtree containing three given vertices---and showed how to solve it in polynomial time via the so-called extended strip decompositions, in order to obtain a~polynomial algorithm for \textsc{$K_{2,3}$-ISC}~\cite{DBLP:journals/combinatorica/ChudnovskyS10}.
The first graphs $H$ for which \textsc{$H$-ISC} is NP-complete came from~\cite{LevequeLMT09} where several examples are given: notably the complete graph $K_5$ and some trees, among other graphs.
In the same paper, some other examples of tractable \textsc{$H$-ISC} were given, all relying on the polynomial-time algorithm for \textsc{Three-In-A-Tree}.
There are also some ad hoc algorithms for \textsc{$H$-ISC} when $H$ is the \emph{net} (i.e., the graph obtained by adding a pendant neighbor to each vertex of a~triangle)~\cite{DBLP:journals/jct/ChudnovskyST13}, when $H$ is $K_4$~\cite{DBLP:journals/jgt/Le19}, or when $H$ is the disjoint union of a~fixed number of triangles~\cite{DBLP:journals/jctb/NguyenSS24}.

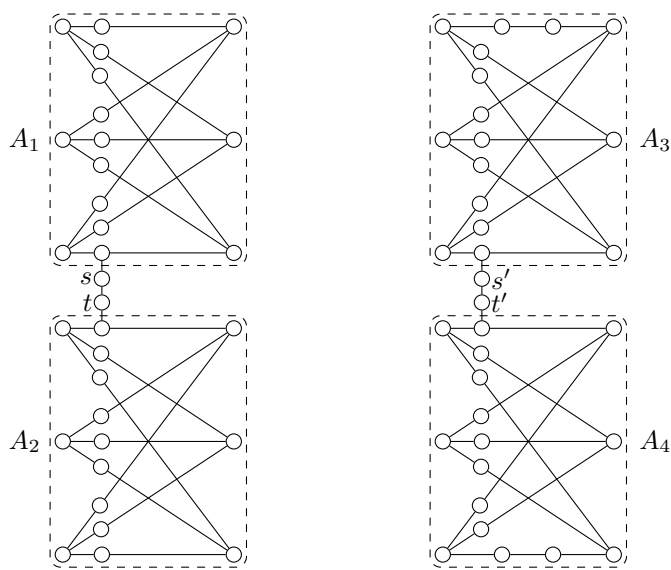
\begin{figure}[h!]
  \centering
  \begin{tikzpicture}[vertex/.style={draw,circle,inner sep=0.05cm, minimum width=0.2cm}]
    \def\s{1.5}

    \foreach \x/\y/\l in {0/0/1, 0/4/2}{
    \begin{scope}[xshift=\x cm, yshift = \y cm] 
      
    \foreach \i in {1,2,3}{
      \node[vertex] (u\l\i) at (0,\i * \s) {} ;
      \node[vertex] (v\l\i) at (1.5 * \s,\i * \s) {} ;
    }
    \node[draw,thin,dashed,rounded corners, inner sep=0.06cm, fit=(u\l1) (v\l3)] {} ;
    
    \foreach \i in {1,2,3}{
      \foreach \j in {1,2,3}{
        \path (u\l\i) to node[vertex, pos=0.2] (w\l\i\j) {} (v\l\j) ;
        \draw (u\l\i) -- (w\l\i\j) -- (v\l\j) ;
      }
    }

    \end{scope}
    }

    \foreach \x/\y/\l in {5/0/3, 5/4/4}{
    \begin{scope}[xshift=\x cm, yshift = \y cm] 
      
    \foreach \i in {1,2,3}{
      \node[vertex] (u\l\i) at (0,\i * \s) {} ;
      \node[vertex] (v\l\i) at (1.5 * \s,\i * \s) {} ;
    }
    \node[draw,thin,dashed,rounded corners, inner sep=0.06cm, fit=(u\l1) (v\l3)] {} ;
    
    \foreach \i/\j in {1/2,1/3,2/1,2/2,2/3,3/1,3/2}{
      \path (u\l\i) to node[vertex, pos=0.2] (w\l\i\j) {} (v\l\j) ;
      \draw (u\l\i) -- (w\l\i\j) -- (v\l\j) ;
    }

    \end{scope}
    }
    \path (u43) to node[vertex, pos=0.33] (w433) {} (v43) ;
    \path (u43) to node[vertex, pos=0.66] (w433b) {} (v43) ;
    \draw (u43) -- (w433) -- (w433b) -- (v43) ;

    \path (u41) to node[vertex, pos=0.2] (w411) {} (v41) ;
    \draw (u41) -- (w411) -- (v41) ;

    \path (u31) to node[vertex, pos=0.33] (w311) {} (v31) ;
    \path (u31) to node[vertex, pos=0.66] (w311b) {} (v31) ;
    \draw (u31) -- (w311) -- (w311b) -- (v31) ;

    \path (u33) to node[vertex, pos=0.2] (w333) {} (v33) ;
    \draw (u33) -- (w333) -- (v33) ;

    % 2 edges
    \path (w133) to node[vertex, pos=0.3] (t) {} (w211) ;
    \path (w133) to node[vertex, pos=0.7] (s) {} (w211) ;
    \draw (w133) -- (t) -- (s)-- (w211) ;

    \path (w333) to node[vertex, pos=0.3] (tp) {} (w411) ;
    \path (w333) to node[vertex, pos=0.7] (sp) {} (w411) ;
    \draw (w333) -- (tp) -- (sp) -- (w411) ;

    \node[left] at (s) {$s$};
    \node[left] at (t) {$t$};
    \node[right] at (sp) {$s'$};
    \node[right] at (tp) {$t'$};

    \node at (-0.5,7) {$A_1$} ;
    \node at (-0.5,2 * \s) {$A_2$} ;

     \node at (7.8,7) {$A_3$} ;
    \node at (7.8,2 * \s) {$A_4$} ;
    
  \end{tikzpicture}
  \caption{The subcubic graph~$H$.
   Both $H[A_1]$ and $H[A_2]$ are the 1-subdivision of $K_{3,3}$.
  Both $H[A_3]$ and $H[A_4]$ are obtained from $K_{3,3}$ by subdividing every edge once but one edge that is subdivided twice.
  The vertices of $H$ which are not in $\bigcup_{i \in [4]} A_i$ are labeled $s, t, s', t'$.
  In total $H$ has 66 vertices.}
  \label{fig:H}
\end{figure}

Despite this line of work, no subcubic graph $H$ was known to make \textsc{$H$-ISC} NP-hard.
Actually, Chudnovsky, Seymour, and the fourth author~\cite{DBLP:journals/jct/ChudnovskyST13} and Le~\cite{DBLP:journals/jgt/Le19} asked whether there is a~polynomial-time algorithm for \textsc{$H$-ISC} for any subcubic graph~$H$.
As a~consequence of~\cref{thm:hardness-I2DP}, we answer this question by the negative by exhibiting a~subcubic graph $H$ (the graph of~\cref{fig:H}) for which \textsc{$H$-ISC} is NP-hard.

\begin{theorem}\label{thm:hardness-subcubic-subd}
  \textsc{$H$-Induced Subdivision Containment} is NP-complete for the subcubic graph $H$ of~\cref{fig:H}.
\end{theorem}

\Cref{thm:hardness-subcubic-subd} holds within subgraphs of a~constant power of bounded-degree planar graphs that are string graphs plus a~constant number of (bounded-degree) apices, which are graphs of bounded twin-width~\cite{twin-width1,twin-width2} excluding a~fixed induced minor.
In other words, \cref{thm:hardness-subcubic-subd} holds in the graphs for which~\cref{thm:hardness-I2DP} holds augmented by a~constant number of vertices of bounded degree.

\medskip

\Cref{thm:hardness-I2DP} also has consequences for the detection of induced minors, topic that we now briefly survey.  
Let \textsc{$H$-Induced Minor Containment} (\textsc{$H$-IMC} for short) input a~graph~$G$ and ask whether $H$ is an induced minor of~$G$.
It was first shown by Fellows et al.~\cite{Fellows95} that \textsc{$H$-IMC} can be NP-hard for a~fixed graph~$H$ (unlike the minor containment).
The latter graph $H$ was not planar, which prompted the authors to ask if there is a~planar graph~$H$ for which \textsc{$H$-IMC} is NP-hard, and whether \textsc{$H$-IMC} is always polynomial-time solvable when $H$ is a~tree.
Recently, Korhonen and Lokshtanov~\cite{KorhonenL23} answered both questions by showing that \textsc{$H$-IMC} is NP-hard for some fixed tree~$H$ (whose number of vertices is not explicit, and estimated to be larger than $2^{300}$ in~\cite{Dallard25}).
Dallard et al.~\cite{Dallard24} show that \textsc{$K_{2,3}$-IMC} can be solved in polynomial time.
Finding the disjoint union of a~constant number of triangles as an induced minor or as an induced subdivision is all the same, so for any natural~$t$, \textsc{$tK_3$-IMC} is also in P~\cite{DBLP:journals/jctb/NguyenSS24}.
For a~more complete survey on the complexity of \textsc{$H$-IMC}, we refer the reader to the introduction of~\cite{Dallard25}; the paper provides some polynomial-time algorithms for three infinite families of graphs~$H$, and notes that \textsc{$H$-IMC} is in P for every graph $H$ on at~most 5 vertices, except for three remaining open cases.

As a~similar consequence to \Cref{cor:tww}, we show a~$2^{\Omega(\sqrt{n})}$ lower bound for \textsc{$H$-IMC} on $n$-vertex graphs, under the ETH.
This can be put in perspective with a~$2^{\tilde{O}(n^{2/3})}$-time algorithm for \textsc{$H$-IMC} when every edge of~$H$ is incident to a~vertex of degree at~most~2~\cite{KorhonenL23}.
The authors ask if the mere NP-hardness of \textsc{$H$-IMC} can be shown for some graph $H$ with this property. 
Actually by subdividing eight edges in the graph of~\cref{fig:H}, we obtain a~graph $H'$ satisfying the property, and for which we can show the same lower bound.

\begin{theorem}\label{thm:hardness-subcubic-ind-minor}
  There is a~subcubic graph $H'$ such that every edge of $H'$ is incident to a~vertex of~degree~2, and \textsc{$H'$-Induced Minor Containment} is NP-complete, and requires time $2^{\Omega(\sqrt n)}$ on $n$-vertex graphs, unless the ETH fails.
\end{theorem}

We observe that it is not too difficult to show~\cref{thm:hardness-subcubic-subd,thm:hardness-subcubic-ind-minor} with \emph{connected} graphs $H$ and $H'$ (having the extra properties of their respective statement).
As this complicates a~bit their proofs without being a~significantly stronger result, we opted against doing it explicitly.

\paragraph*{Open questions}

We suggest the following open questions, which come more or less directly from our work and the literature.
In light of the surveyed polynomial algorithms applying more generally to \textsc{Induced $k$-Disjoint Paths} and the hardness proofs applying more generally to \textsc{Induced Disjoint $S$--$T$ Paths} with $|S|=|T|=k$, we wonder if there is a hereditary graph class in which \textsc{Induced $k$-Disjoint Paths} is NP-complete but \textsc{Induced Disjoint $S$--$T$ Paths} with $|S|=|T|=k$ is polynomial-time solvable; the case $k=2$ is of particular interest.  

The string graphs that our proof of~\cref{thm:hardness-I2DP} produces do not seem to be 1-string graphs (i.e., realizable with strings every pair of which intersects at~most~once).
We leave open the complexity of \textsc{Induced $k$-Disjoint Paths} in 1-string graphs, and also in segment intersection graphs (a~further restriction to 1-string graphs).

A~natural conjecture stems from our paper and existing algorithms, under P $\neq$ NP.
\begin{conjecture}\label{conj:subcubic-subd-planar}
 For any subcubic graph $H$, \textsc{$H$-ISC} is in P if and only if $H$ is planar.
\end{conjecture}

\section{Preliminaries}

If $i$ is a~positive integer, we denote by $[i]$ the set of integers $\{1,2,\ldots,i\}$.

\subsection{Subgraphs, induced subgraphs, neighborhoods, and some graphs}\label{sec:graph-def}

We denote by $V(G)$ and $E(G)$ the set of vertices and edges of a graph $G$, respectively.
A~graph $H$ is a~\emph{subgraph} of a~graph $G$ if $H$ can be obtained from $G$ by vertex and edge deletions.
Graph~$H$ is an~\emph{induced subgraph} of $G$ if $H$ is obtained from $G$ by vertex deletions only.
A~graph $G$ is \emph{$H$-free} if $G$ does not contain $H$ as an induced subgraph.
For $S \subseteq V(G)$, the \emph{subgraph of $G$ induced by $S$}, denoted $G[S]$, is obtained by removing from $G$ all the vertices that are not in $S$ (together with their incident edges).
Then $G-S$ is a short-hand for $G[V(G)\setminus S]$.
A collection of paths $P_1, \ldots, P_h$ in a~graph $G$ is said \emph{mutually induced} if $G[\bigcup_{i \in [h]} V(P_i)]$ is exactly the disjoint union of paths $P_1, \ldots, P_h$.

A~set $X \subseteq V(G)$ is connected (in $G$) if $G[X]$ has a~single connected component.
A~vertex whose removal increases the number of connected components is called a~\emph{cutvertex}.
Similarly, a~\emph{bridge} is an edge whose removal increases the number of connected components.
We denote by $G^r$ the \emph{$r$-th power of $G$}, that is, the graph with vertex set $V(G)$ and an edge between any two vertices at (shortest-path) distance at~most~$r$ in~$G$.
A~\emph{constant power} of~$G$ is $G^r$ for some constant~$r$. 

A~\emph{subdivision} of a~graph $G$ is any graph obtained from $G$ by replacing each edge of~$G$ by a~path of at~least~one edge.
In a~subdivision of~$G$ the vertices of~$G$ are called \emph{branching vertices}, and the created vertices are called \emph{subdivision vertices}.
The \emph{$s$-subdivision} of $G$ is the graph obtained from $G$ by replacing each edge of~$G$ by a~path of $s+1$ edges.
 
We denote by $N_G(v)$ and $N_G[v]$, the open, respectively closed, neighborhood of $v$ in $G$.
For $S \subseteq V(G)$, we set $N_G(S) := (\bigcup_{v \in S}N_G(v)) \setminus S$ and $N_G[S] := N_G(S) \cup S$.
The \emph{degree} $d_G(v)$ of a~vertex $v \in V(G)$ is the cardinality of $N_G(v)$, and the \emph{maximum degree} of $G$ is defined as $\max_{v \in V(G)} d_G(v)$.
A~\emph{subcubic graph} is a~graph of maximum degree at most~3.

The \emph{$t$-clique}, denoted by \emph{$K_t$}, is obtained by making adjacent every pair of two distinct vertices among $t$ vertices, and the \emph{biclique $K_{t,t}$} with bipartition $(A,B)$ such that $|A|=|B|=t$ is obtained by making every vertex of~$A$ adjacent to every vertex of~$B$.  
A~\emph{theta} is any subdivision of $K_{2,3}$.
A~\emph{wheel} is any graph obtained by adding to a~cycle of length at~least~4, a~vertex with at~least three neighbors on the cycle.

A~\emph{string graph} is the intersection graph of some collection of (non-self-intersecting) curves in the plane (usually called strings), or equivalently the intersection graph of a~collection of connected sets of some planar graph.
The collection of strings is called \emph{string representation}.
We may see the string representation as a~planar diagram with one vertex at each string endpoint and at each intersection of two strings.
For instance, any string representation defines an infinite face (the infinite face of this planar diagram).

It is known (and easy to see) that the 1-subdivision of any non-planar graph is not a~string graph~\cite{Sinden66}.

\subsection{Induced subdivisions and induced minors}\label{sec:induced-subd-min}

A~graph $H$ is an induced subdivision of a~graph~$G$ if a~subdivision of~$H$ is isomorphic to an induced subgraph of~$G$.
An \emph{induced subdivision model} of $H$ in $G$ is given by an injective map $\phi: V(H) \to V(G)$ and a~collection of paths $(P_e)_{e \in E(H)}$ in $G$ such that for every $uv \in E(H)$, $P_{uv}$ is a~$\phi(u)$--$\phi(v)$ path, and $G[\bigcup_{e \in E(H)} V(P_e)]$ has no more edges than the paths $(P_e)_{e \in E(H)}$ themselves. 

An~\emph{induced minor model} of $H$ in~$G$ is a~collection $\mathcal M := \{X_1, \ldots, X_{|V(H)|}\}$ of pairwise-disjoint connected subsets of $V(G)$, called \emph{branch sets}, together with a~bijective map $\phi: V(H) \to \mathcal M$ such that $uv \in E(H)$ if and only if there is at~least one edge in $G$ between $\phi(u)$ and $\phi(v)$.
In which case, we may say that the branch sets $\phi(u)$ and $\phi(v)$ are \emph{adjacent}.
We then say that $H$ is an \emph{induced minor} of $G$ (or otherwise that $G$ is $H$-induced-minor-free).
Or equivalently, $H$ can be obtained from $G$ after a~series of~vertex deletions and edge contractions.

An induced minor model $(\{X_1, \ldots, X_h\}, \phi)$ of an $h$-vertex graph $H$ is \emph{minimal} if for every $X'_1 \subseteq X_1, \ldots, X'_h \subseteq X_h$, the fact that $(\{X'_1, \ldots, X'_h\}, \phi')$ is an induced minor model of~$H$ with $\phi'(u) = X'_i \Leftrightarrow \phi(u) = X_i$ for every $u \in V(H)$, implies that for every $i \in [h]$, $X'_i = X_i$.

With the second given definition of string graphs (see end of~\cref{sec:graph-def}), it is easy to see that the class of string graphs is closed under taking induced minors.
Thus no string graph admits the 1-subdivision of a~non-planar graph as an induced minor. 

\subsection{Useful facts on twin-width}

As we only mention twin-width in side remarks, we refrain from giving a~definition.
We list the theorems useful in~\cref{sec:i2dp}.
This can be read in a~black-box fashion.

It was first proven in~\cite{twin-width1} that the class of planar graphs has bounded twin-width.
The current best upper bound is~8~\cite{HlinenyJ23}.

\begin{theorem}[Theorem 6.3 in \cite{twin-width1}, \cite{HlinenyJ23}]\label{thm:planar-tww}
  Planar graphs have bounded twin-width, more precisely upper bounded by~8.
\end{theorem}

The constant powers of bounded twin-width graphs have bounded twin-width, as a~special case of so-called first-order transductions.

\begin{theorem}[Theorem 8.1 in \cite{twin-width1}]\label{thm:power-tww}
  There is a~function $f$ such that for any graph $G$ of twin-width~$d$ and for any positive integer~$r$, $G^r$ has twin-width at~most~$f(d,r)$.
\end{theorem}

Among weakly sparse classes (excluding a~$K_{t,t}$ as a~subgraph), the subgraph closure of any class of bounded twin-width has bounded twin-width.  

\begin{theorem}[\cite{twin-width2}]\label{thm:ws-subgraph-closure}
  There is a~function $g$ such that for any graph $G$ of twin-width~$d$ excluding~$K_{t,t}$ as a~subgraph (or in particular, of maximum degree at~most~$t-1$), then any subgraph of~$G$ has twin-width at~most~$g(d,t)$. 
\end{theorem}

\section{Hardness of \textsc{Induced 2-Disjoint Paths} in string graphs}\label{sec:i2dp}

In this section, we show that \textsc{Induced 2-Disjoint Paths} is NP-complete in (a~proper subclass of) string graphs.

\begin{reptheorem}{thm:hardness-I2DP}
  \textsc{Induced 2-Disjoint Paths} is NP-complete in string graphs that are subgraphs of a~constant power of bounded-degree planar graphs.
\end{reptheorem}

\begin{proof}
  We reduce from the NP-complete problem \textsc{Clause-Linked Planar E3-Occ 3-SAT}, with variables $x_1, \ldots, x_n$ and clauses $c_1, \ldots, c_m$ where every clause $c_j$ is on at~most three variables, each variable appears three times, and the variable-clause incidence graph is planar even when the cycle $c_1c_2 \ldots c_mc_1$ is added.
  This problem has been proven NP-complete by Fellows et al.~\cite{Fellows95}, even in a~restricted form with some additional constraint on the clauses, but we will not need this restriction.
  We also do not absolutely need that every variable appears exactly three times; \emph{at~most three times} is good enough.
  To get a~unique representation of the variable gadget, we nevertheless assume that every variable has exactly three occurrences: two positive and one negative, or two negative and one positive.
  Indeed variables appearing only positively (or only negatively) can be set to true (or to false).
  This means that we need to allow clauses on two variables, as \textsc{E3-Occ E3-SAT} is trivial (all its instances are satisfiable).
  We refer the reader to~\cite{Tippenhauer16} for the complexity of variants of~\textsc{Planar 3-SAT}.
  
  We present the variable gadgets, clause gadgets, and variable-clause incidences, which assembled together form the graph $G$ input to~\textsc{Induced 2-Disjoint Paths}.
  The reader can also refer to \cref{fig:clause-gadget-and-neig} where these three elements are depicted.

  \medskip

  \textbf{Variable gadget.}
  For each variable $x_i$, we now describe the variable gadget $\mathcal G(x_i)$.
  For each clause $c_j$ containing $x_i$, we add to $\mathcal G(x_i)$ two vertices $w_j^N(x_i), w_j^S(x_i)$ if $x_i$ appears positively in~$c_j$, and we add two vertices $w_j^N(\neg x_i), w_j^S(\neg x_i)$ if, instead, $x_i$ appears negatively in~$c_j$.
  We finally turn $\mathcal G(x_i)$ into an induced biclique by adding every edge between pairs of vertices $w_j^d(x_i), w_{j'}^{d'}(\neg x_i)$ for $d, d' \in \{N,S\}$ and $j, j'$ may be equal or distinct.
  Observe that every variable gadget is isomorphic to the bipartite complete graph~$K_{2,4}$.

 \medskip

 \textbf{Clause gadget.}
 We describe the clause gadget $\mathcal G(c_j)$ for each clause $c_j$.
 Let $\ell_1, \ell_2, \ell_3$ be the three literals of~$c_j$ (or $\ell_1, \ell_2$ if $c_j$ is a~2-clause).
 The gadget $\mathcal G(c_j)$ has 16 vertices (or 12 if $c_j$ is a~2-clause): four \emph{entry points} $u^N_j, u^S_j, u^N_{j+1}, u^S_{j+1}$, and four vertices $v_j^{NW}(\ell_a), v_j^{NE}(\ell_a), v_j^{SW}(\ell_a), v_j^{SE}(\ell_a)$ for each $a \in \{1,2,3\}$ (for each $a \in \{1,2\}$).
 We add an edge between a~pair of vertices $v_j^d(\ell_a)$, $v_j^{d'}(\ell_{a'})$, with $d, d' \in \{NW,NE,SW,SE\}$, whenever $a \neq a'$.
 Thus those 12 vertices (8 vertices) form a~tripartite complete graph $K_{4,4,4}$ (bipartite complete graph $K_{4,4}$).
 For every $d \in \{N,S\}$ and $a \in \{1,2,3\}$ ($a \in \{1,2\}$), we also add the six (four) edges $u^d_j v_j^{dW}(\ell_a)$, and the six (four) edges $v_j^{dE}(\ell_a) u^d_{j+1}$.
 This finishes the description of~$\mathcal G(c_j)$.
 Note that $\mathcal G(c_j)$ and $\mathcal G(c_{j+1})$ share two vertices: $u^N_{j+1}$ and $u^S_{j+1}$.

 \medskip

 \textbf{Variable-clause incidence.}
 Finally for every clause $c_j$ and literal $\ell$ in~$c_j$, and each $d \in \{N,S\}$, we add the edges $v_j^{dW}(\ell)w_j^d(\ell)$ and $w_j^{dE}(\ell)v_j^{d}(\ell)$.

 \medskip
 
 The graph~$G$, input of \textsc{Induced 2-Disjoint Paths}, is obtained by having a~gadget for each variable and each clause, and adding the edges as described in the previous sentence. 
 We set the first terminal pair to $u^N_1, u^N_{m+1}$ and the second terminal pair to $u^S_1, u^S_{m+1}$.
 This finishes the construction; see~\cref{fig:clause-gadget-and-neig}.  
 \begin{figure}[h!]
   \centering
   \begin{tikzpicture}[vertex/.style={draw,circle}]
     \def\s{4}
     \def\ss{1.75}
     \def\v{0.4}
     % literals K_{4,4,4}
     \foreach \l/\i in {{x_{j_1}}/0,{\neg x_{j_2}}/1,{x_{j_3}}/2}{
       \foreach \d/\x/\y/\z in {NW/0/\ss/-\v,NE/\ss/\ss/-\v,SW/0/0/\v,SE/\ss/0/\v}{
         \node[vertex] (\d\i) at (\i * \s + \x, \y) {} ;
         \node (lab\d\i) at (\i * \s + \x, \y + \z) {\textcolor{white}{\tiny{$v^{\d}_j(\l)$}}} ;
       }
       \node[draw, thick, rounded corners, fit={(NW\i) (SE\i) (labNW\i) (labSE\i)}] (lit\i) {} ;
     }

     % entrypoints
     \foreach \j/\ji/\d/\x/\y/\z in {j/j/N/-1.25/\ss/-\v,j/j/S/-1.25/0/\v,{j+1}/k/N/11/\ss/-\v,{j+1}/k/S/11/0/\v}{
       \node[vertex] (u\d\ji) at (\x, \y) {} ;
       \node (labu\d\ji) at (\x, \y + \z) {\tiny{$u^\d_{\j}$}} ;
     }

     % clause gadget edges
    \draw[very thick] (lit0) -- (lit1) -- (lit2) to [bend left = 29] (lit0) ;
    \foreach \x/\y/\b/\c in {
      uNj/NW0/0/red,uNj/NW1/20/blue,uNj/NW2/22/green!75!black, uNk/NE2/0/green!75!black,uNk/NE1/-20/blue,uNk/NE0/-22/red,
      uSj/SW0/0/red,uSj/SW1/-20/blue,uSj/SW2/-22/green!75!black, uSk/SE2/0/green!75!black,uSk/SE1/20/blue,uSk/SE0/22/red}{
      \draw[thick, \c] (\x) to [bend left=\b] (\y) ;
    }

    \draw[dashed, rounded corners] (-1.6,-1.1) -- (-1.6,2.85) -- (11.5, 2.85) -- (11.5, -1.1) -- cycle ;
    \node at (-1, -0.75) {$\mathcal G(c_j)$} ;
    
    % variable gadgets
    \foreach \l/\x/\y/\d/\lit in {
      wNj0/{\ss/2}/4.3/N/{x_{j_1}},wSj0/{\ss/2}/3.6/S/{x_{j_1}},
      wNj2/{2 * \s + \ss/2}/-2/N/{x_{j_3}},wSj2/{2 * \s + \ss/2}/-2.7/S/{x_{j_3}}}{
      \node[vertex] (\l) at (\x, \y) {} ;
      \node (lab\l) at (\x - 1, \y) {\tiny{$w_j^\d(\lit)$}} ;
    }
    \foreach \l/\x/\y/\d/\lit in {
      wNj1/{\s+\ss/2}/4.3/N/{\neg x_{j_2}},wSj1/{\s+\ss/2}/3.6/S/{\neg x_{j_2}}
    }{
      \node[vertex] (\l) at (\x, \y) {} ;
      \node (lab\l) at (\x + 1, \y) {\tiny{$w_j^\d(\lit)$}} ;
    }
    \foreach \l/\x/\y/\d/\lit/\j/\z in {
      wNjp/{\ss}/5.8/N/{x_{j_1}}/{j'}/1,wSjp/{\ss}/5.1/S/{x_{j_1}}/{j'}/1,
      wNjpp/0/5.8/N/{\neg x_{j_1}}/{j''}/-1,wSjpp/0/5.1/S/{\neg x_{j_1}}/{j''}/-1}{
      \node[vertex] (\l) at (\x, \y) {} ;
      \node (lab\l) at (\x + \z, \y) {\tiny{$w_{\j}^\d(\lit)$}} ;
    }
    \foreach \f/\l in {{(wNj0) (wSj0)}/wj0, {(wNjp) (wSjp)}/wjp, {(wNjpp) (wSjpp)}/wjpp}{
      \node[draw, thick, rounded corners, fit={\f}] (\l) {} ;
    }
    \draw[very thick] (wj0) -- (wjpp) -- (wjp) ;

    \draw[dashed, rounded corners] (-1.9,3) -- (-1.9,6.2) -- (3.6, 6.2) -- (3.6, 3) -- cycle ;
    \node at (-1.2,3.35) {$\mathcal G(x_{j_1})$} ;

    % clause-literal edges
    \foreach \x/\y/\b/\c in {
      NW0/wNj0/0/red,NE0/wNj0/0/red, NW1/wNj1/0/blue,NE1/wNj1/0/blue, NW2/wNj2/0/green!75!black,NE2/wNj2/0/green!75!black,
      SW0/wSj0/0/red,SE0/wSj0/0/red, SW1/wSj1/0/blue,SE1/wSj1/0/blue, SW2/wSj2/0/green!75!black,SE2/wSj2/0/green!75!black}{
      \draw[thick, \c] (\x) to [bend left=\b] (\y) ;
    }

     \foreach \l/\i in {{x_{j_1}}/0,{\neg x_{j_2}}/1,{x_{j_3}}/2}{
       \foreach \d/\x/\y/\z in {NW/0/\ss/-\v,NE/\ss/\ss/-\v,SW/0/0/\v,SE/\ss/0/\v}{
         \node (lab\d\i) at (\i * \s + \x, \y + \z) {\tiny{$v^{\d}_j(\l)$}} ;
       }
     }
   \end{tikzpicture}
   \caption{The clause gadget $\mathcal G(c_j)$ with $c_j = x_{j_1} \lor \neg x_{j_2} \lor x_{j_3}$, variables $x_{j_1}, x_{j_2}$ in one face defined by cycle $c_1c_2 \ldots c_mc_1$ (above), and $x_{j_3}$ in the other face (below).
     We also drew the variable gadget $\mathcal G(x_{j_1})$ when $x_{j_1}$ appears positively in $c_j$ and $c_{j'}$, and negatively in $c_{j''}$. 
     Edges linking two rounded boxes represent bicliques. %, where individual edges are actually drawn as straight-line segments.
   }
   \label{fig:clause-gadget-and-neig}
 \end{figure}
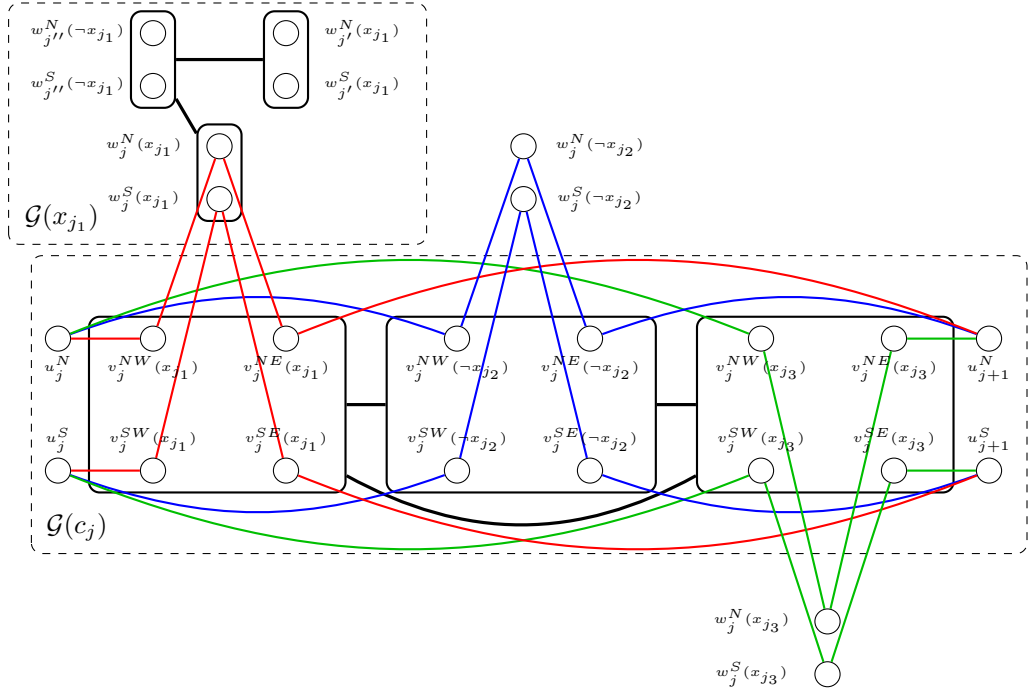
  The next two lemmas show the correctness of the reduction.

  \begin{claim}
    If $c_1 \land \ldots \land c_m$ is satisfiable, then $G$ admits two $u^N_1$--$u^N_{m+1}$ and $u^S_1$--$u^S_{m+1}$ mutually induced paths.
  \end{claim}
 \begin{proofofclaim}
 We fix a~truth assignment $\ell_1, \ldots, \ell_n$ of the variables satisfying $c_1 \land \ldots \land c_m$, with $\ell_i \in \{x_i, \neg x_i\}$ for every $i \in [n]$.
 For each clause $c_j$, we choose any literal $\ell(c_j)$ of $c_j$ such that $\ell(c_j) = \ell_i$ for some $i \in [n]$. 
 For each $d \in \{N,S\}$, we build the path~$P^d$:
 \[u^d_1, v_1^{dW}(\ell(c_1)), w_1^d(\ell(c_1)), v_1^{dE}(\ell(c_1)),~u^d_2, v_2^{dW}(\ell(c_2)), w_2^d(\ell(c_2)), v_2^{dE}(\ell(c_2)),~\ldots,\]
 \[u^d_m, v_m^{dW}(\ell(c_m)), w_m^d(\ell(c_m)), v_m^{dE}(\ell(c_m)), u^d_{m+1}.\]
 Note that $P^N$ is a~$u^N_1$--$u^N_{m+1}$ path in~$G$, $P^S$ is a~$u^S_1$--$u^S_{m+1}$ path, and $P^N, P^S$ are vertex-disjoint.
 We claim there is no chord between $P^N$ and $P^S$.
 This is essentially because
 \begin{compactitem}
 \item within every $\mathcal G(c_j)$, $V(P^N)$ and $V(P^S)$ induce a~$4K_2$ comprising four edges of $P^N \uplus P^S$,
 \item apart from edges incident to entry points, there is no edge between two clause gadgets,  
 \item $P^N$ and $P^S$ enter each variable gadget on the same side of the induced biclique,
 \end{compactitem}
 which concludes the proof of the claim.
 \end{proofofclaim}

 \begin{claim}\label{clm:paths-to-sat}
   If $G$ admits two $u^N_1$--$u^N_{m+1}$ and $u^S_1$--$u^S_{m+1}$ mutually induced paths, then $c_1 \land \ldots \land c_m$ is satisfiable.
 \end{claim}
 \begin{proofofclaim}
 Let $P^N, P^S$ be two mutually induced paths in $G$ such that $P^d$ is a~$u^d_1$--$u^d_{m+1}$ path for each $d \in \{N,S\}$.
 We think of $P^d$ as going from $u^d_1$, its start, to $u^d_{m+1}$, its end.
 We first show the following invariant.
 For every $j \in [m]$, there exists a~literal $\ell$ of~$c_j$ such that for each $d \in \{N,S\}$, the path $P^d$ goes from $u^d_j$ to $u^d_{j+1}$ via the 4-edge subpath $u^d_j, v_j^{dW}(\ell), w_j^d(\ell), v_j^{dE}(\ell), u^d_{j+1}$.
 
 Fix some $j \in [m]$, and assume that the invariant holds for every $j' \in [j-1]$ (with $[0] = \emptyset$).
 The vertex following $u^N_j$ along $P^N$ cannot be in $\mathcal G(c_{j-1})$ as either $j=1$ (and this gadget does not exist) or it would create a chord between $P^N$ and $P^S$, by the induction hypothesis.
 Thus it has to be $v_j^{NW}(\ell)$ for some literal $\ell$ of~$c_j$.
 To avoid a chord between $P^N$ and $P^S$, the vertex following $u^S_j$ along $P^S$ has to be $v_j^{SW}(\ell)$.
 Then the next vertex along $P^N$ (resp.~$P^S$) has to be $w_j^N(\ell)$ (resp.~$w_j^S(\ell)$).
 As the variable gadget of literal~$\ell$ is an induced biclique, the next vertices have to be back to the clause gadget $\mathcal G(c_j)$: $v_j^{NE}(\ell)$ along $P^N$, and $v_j^{SE}(\ell)$ along $P^S$.
 Finally the next vertices have to be $u^N_{j+1}$ along $P^N$, and $u^S_{j+1}$ along $P^S$.
 This completes the proof that the invariant holds for every $j \in [m]$. 

 Let $\mathcal A$ be the truth assignment setting $x_i$ to true if $P^N$ (resp.~$P^S$) does not contain any vertex $w^N_j(\neg x_i)$ (resp.~$w^S_j(\neg x_i)$), and to false otherwise.
 As $P^N, P^S$ are mutually induced, in the latter case $P^N$ (resp.~$P^S$) does not contain any vertex $w^N_j(x_i)$ (resp.~$w^S_j(x_i)$).
 Thus, for every clause $c_j$, the vertex $v_j^{NW}(\ell) \in P^N$ is such that literal $\ell$ is satisfied by~$\mathcal A$.
 Hence $\mathcal A$ is a~satisfying assignment. 
 \end{proofofclaim}

 Observe that the invariant in the first paragraph of the proof of~\cref{clm:paths-to-sat} still holds under the weaker assumption that the endpoints of $P^N, P^S$ are in $\{u^N_{m+1}, u^S_{m+1}\}$ (but are unspecified).
 Therefore finding two mutually induced paths between $\{u^N_1, u^S_1\}$ and $\{u^N_{m+1}, u^S_{m+1}\}$ in $G$ (the induced flow problem) is equivalent to the induced linkage problem. 

 We now show that $G$ is as advertised by the theorem statement.
 We start by giving a~string representation for~$G$.
 
 \begin{claim}\label{clm:G-string-graph}
   $G$ is a~string graph.
 \end{claim}
 \begin{proofofclaim}
 First, we draw a~planar embedding $\mathcal P$ of the variable-clause incidence graph augmented with the cycle $c_1c_2 \ldots c_mc_1$.
 We refer to one face delimited by this cycle as the \emph{upper face} (drawn in the figures ``above'' the path $c_1c_2 \ldots c_m$), and the other face as the \emph{lower face}. 
 Second, for every clause~$c_j$, draw $\mathcal G(c_j)$, and the six (or four, if $c_j$ is a~2-clause) vertices in variable gadgets that are adjacent to $\mathcal G(c_j)$, as the intersection graph of strings such that
 \begin{compactitem}
 \item the strings of $u^N_j, u^S_j, u^N_{j+1}, u^S_{j+1}$ protrude in the infinite face,\footnote{In all these items, the faces are those of the planar diagram  of the string representation of~$\mathcal G(c_j)$, except the upper and lower faces, which are defined above.}
 \item as well as $w^N_j(\ell)$ (resp.~$w^S_j(\ell)$) if $\ell$ is a~literal of~$c_j$ whose variable is drawn in the upper face (resp.~lower face) in~$\mathcal P$,
 \item these strings appear along the infinite face in the order prescribed by~$\mathcal P$, and
 \item for each literal $\ell$ of~$c_j$, there is a~face that contains a~substring of $w^N_j(\ell)$ and a~substring of~$w^S_j(\ell)$, one of which bounding the infinite face (as requested by the second item).
 \end{compactitem}
 Such a~string representation is given in~\cref{fig:string-graph-rep}.

 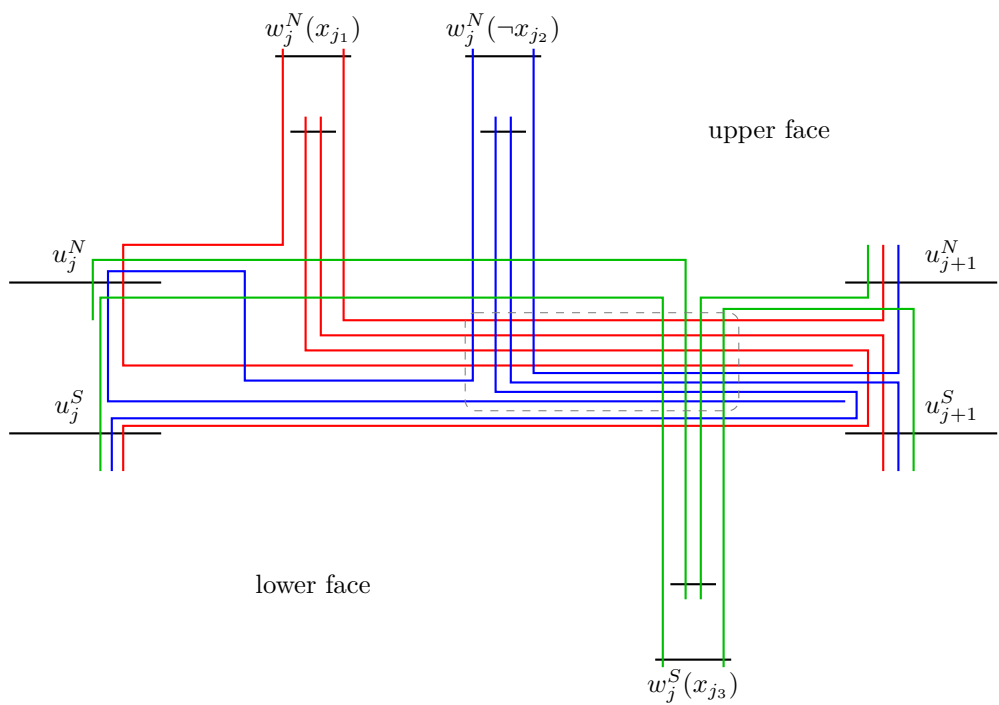
\begin{figure}[h!]
   \centering
   \begin{tikzpicture}[string/.style={thick}]
     % short strings
     \foreach \i/\j/\l in {0/1/{u^N_j},0/-1/{u^S_j}}{ 
       \draw[string] (\i-1,\j) to node[pos=0.4, above] {$\l$} (\i + 1,\j) ;
     }
     \foreach \i/\j/\l in {10/1/{u^N_{j+1}},10/-1/{u^S_{j+1}}}{ 
       \draw[string] (\i,\j) to node[pos=0.7, above] {$\l$} (\i + 2,\j) ;
     }
     \foreach \i/\j/\l in {2.5/4/{w^N_j(x_{j_1})}, 5/4/{w^N_j(\neg x_{j_2})}}{ 
       \draw[string] (\i,\j) to node[midway, above] {$\l$} (\i + 1,\j) ;
     }
      \foreach \i/\j/\l in {7.5/-4/{w^S_j(x_{j_3})}}{ 
       \draw[string] (\i,\j) to node[midway, below] {$\l$} (\i + 1,\j) ;
     }
     \foreach \i/\j in {2.5/3, 5/3, 7.5/-3}{ 
       \draw[string] (\i+0.2,\j) -- (\i + 0.8,\j) ;
     }

     % 12 strings
     % red
     \draw[string,red] (10.1, -0.1) -- (0.5, -0.1) -- (0.5,1.5) -- (2.6,1.5) -- (2.6,4.1) ;
     \draw[string,red] (3.4,4.1) -- (3.4,0.5) -- (10.5,0.5) -- (10.5,1.5) ;
     \draw[string,red] (0.5, -1.5) -- (0.5,-0.9)-- (10.3,-0.9) -- (10.3,0.1) -- (2.9,0.1) -- (2.9,3.2) ;
     \draw[string,red] (10.5,-1.5) -- (10.5,0.3) -- (3.1,0.3) -- (3.1,3.2) ;

     % blue
     \draw[string,blue] (10, -0.575) -- (0.3, -0.575) -- (0.3, 1.15) -- (2.1, 1.15) -- (2.1, -0.3) -- (5.1, -0.3) -- (5.1, 4.1) ;
     \draw[string,blue] (0.35, -1.5) -- (0.35, -0.8) -- (10.15, -0.8) -- (10.15, -0.45) -- (5.4, -0.45) -- (5.4, 3.2) ;
     \draw[string,blue] (10.7, -1.5) -- (10.7, -0.325) -- (5.6, -0.325) -- (5.6, 3.2) ;
     \draw[string,blue] (10.7, 1.5) -- (10.7, -0.2) -- (5.9, -0.2) -- (5.9, 4.1) ;

     % green
     \draw[string,green!75!black] (0.1, 0.5) -- (0.1, 1.3) -- (7.9, 1.3) -- (7.9, -3.2) ;
     \draw[string,green!75!black] (0.2, -1.5) -- (0.2, 0.8) -- (7.6, 0.8) -- (7.6, -4.1) ;
     \draw[string,green!75!black] (10.3, 1.5) -- (10.3, 0.8) -- (8.1, 0.8) -- (8.1, -3.2) ;
     \draw[string,green!75!black] (10.9, -1.5) -- (10.9, 0.65) -- (8.4, 0.65) -- (8.4, -4.1) ;

     % all intersections
     \draw[gray, dashed, rounded corners] (5,0.6) -- (8.6,0.6) -- (8.6,-0.7) -- (5,-0.7) -- cycle ;

     % faces
     \node at (9,3) {upper face} ;
     \node at (3,-3) {lower face} ;
   \end{tikzpicture}
   \caption{String representation of the clause gadget of~\cref{fig:clause-gadget-and-neig} satisfying all four items.
     The string intersections for the $3 \cdot 4^2 = 48$ adjacencies of the $K_{4,4,4}$ occur in the dashed box with rounded corners.
     It is also easy to check that no two strings of the same color intersect, so the $K_{4,4,4}$ remains induced.
   We kept ``half'' of the strings of $u^N_j, u^S_j, u^N_{j+1}, u^S_{j+1}$ uncrossed for the representation of clause gadgets $\mathcal G(c_{j-1})$ and $\mathcal G(c_{j+1})$.}
   \label{fig:string-graph-rep}
 \end{figure}

The figure illustrates the case when no three variables of the clause lie on the same face.
However it is easy to mirror the green strings to draw the other case.
For 2-clauses, one can simply remove four strings of the same color (red, blue, or green) from~\cref{fig:string-graph-rep}.

At this point, the string representation of~$G$ has all the desired edges (intersections) except for those of the $K_{2,4}$ in variable gadgets.
It has no extra edge, due to the planarity of~$\mathcal P$.
\Cref{fig:var-gadget-edit} finally shows how to edit the strings $w^N_{\bullet}(\bullet)$ and $w^S_{\bullet}(\bullet)$ to add these edges, without incurring any other edges.

\begin{figure}[h!]
  \centering
  \begin{tikzpicture}[string/.style={thick}]
    \foreach \k in {0,1,2}{
      \begin{scope}[rotate=\k * 120]
    \draw[string] (-0.5,-1) -- (0.5,-1) ;
    \draw[string] (-0.25,-1.5) -- (0.25,-1.5) ;

    \foreach \i/\j in {-0.4/-0.9,0.4/-0.9, -0.2/-0.9,0.2/-0.9}{
      \draw[string, red] (\i, -2) -- (\i, \j) ;
      \draw[string, dashed,  red] (\i, -2) -- (\i, -2.5) ;
    }
    \end{scope}
    }
    \node at (1.1, -1) {$w_j^N(x_i)$} ;
    \node at (1.5, -0.2) {$w_{j'}^N(x_i)$} ;
    \node at (-1.3, -0.2) {$w_{j''}^N(\neg x_i)$} ;

    \node at (3.75,0) {$\longrightarrow$} ;

    \begin{scope}[xshift=7.5cm]
      \foreach \k in {0,1,2}{
      \begin{scope}[rotate=\k * 120]
    \foreach \i/\j in {-0.4/-0.9,0.4/-0.9, -0.2/-0.9,0.2/-0.9}{
      \draw[string, red] (\i, -2) -- (\i, \j) ;
      \draw[string, dashed,  red] (\i, -2) -- (\i, -2.5) ;
    }
    \end{scope}
    }

    \foreach \k in {0,1}{
      \begin{scope}[rotate=\k * 120]
    \draw[string] (-0.5,-1) -- (-0.1,-1) -- (-0.1,-0.1) -- (0.1,-0.1) -- (0.1, -1) -- (0.5,-1) ;
    \draw[string] (-0.25,-1.5) -- (-0.05,-1.5) -- (-0.05,-0.15) -- (0.05,-0.15) -- (0.05,-1.5) -- (0.25,-1.5) ;
    \end{scope}
    }
     
    \begin{scope}[rotate=2 * 120]
    \draw[string] (-0.5,-1) -- (-0.1,-1) -- (-0.6,0.15) -- (0.6,0.15) -- (0.1,-1) -- (0.5,-1) ;
    \draw[string] (-0.25,-1.5) -- (-0.05,-1.5) -- (-0.05,-1) -- (-0.52,0.1) -- (0.52,0.1) -- (0.05,-1) -- (0.05,-1.5) -- (0.25,-1.5) ;
    \end{scope}

    \node at (1.1, -1) {$w_j^N(x_i)$} ;
    \node at (1.5, -0.2) {$w_{j'}^N(x_i)$} ;
    \node at (-1.3, -0.2) {$w_{j''}^N(\neg x_i)$} ;
    \end{scope}
  \end{tikzpicture}
  \caption{How to realize the $K_{2,4}$ of the variable gadget $\mathcal G(x_i)$ without creating any other string intersections.
    The black strings represent vertices of~$\mathcal G(x_i)$, and the red strings, vertices in clause gadgets.
    In this example, $x_i$ is in the upper face, and has two positive occurrences and one negative occurrence.
  }
  \label{fig:var-gadget-edit}
\end{figure}
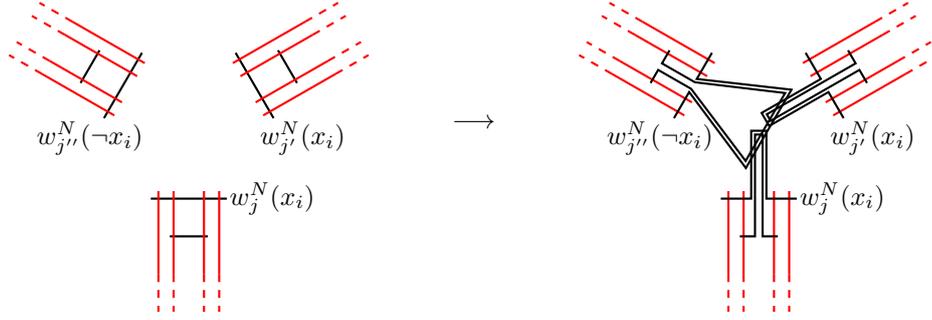
After editing every string of the variable gadgets, we obtain a~string representation for~$G$.
 \end{proofofclaim}

 We finally show the following property of~$G$.
 \begin{claim}\label{clm:bd-tww}
   There is a~subcubic planar graph $H$ such that $G$ is a~subgraph of~$H^{16}$. 
 \end{claim}
 \begin{proofofclaim}
   In the planar drawing $\mathcal P$ (see~proof of~\cref{clm:G-string-graph}), replace every clause vertex by a~path of length 16, and every variable vertex by a~cycle of length 12, in such a~way that the obtained graph $H$ is subcubic, still planar, and the variable-clause incidences are preserved.
   It is then easy to see that $G$ is a~subgraph of~$H^{16}$. 
 \end{proofofclaim}

 This finishes the proof of the main theorem.
\end{proof}

We then derive the following.

\begin{repcorollary}{cor:tww}
  \textsc{Induced Disjoint $S$--$T$ Paths} with $|S|=|T|=2$ is NP-complete in string graphs of bounded maximum degree and twin-width, and requires time $2^{\Omega(\sqrt n)}$ on $n$-vertex such graphs, unless the Exponential-Time Hypothesis fails.
\end{repcorollary}
\begin{proof}
  By~\cref{thm:planar-tww}, planar graphs have bounded twin-width.
  By~\cref{thm:power-tww}, constant powers of planar graphs have bounded twin-width.
  Constant powers of bounded-degree graphs have themselves bounded maximum degree, and in particular, no $K_{t,t}$ subgraph for some constant~$t$.
  Thus by~\cref{thm:ws-subgraph-closure}, the graphs produced by the previous reduction have bounded twin-width.
  Besides, they are bounded-degree string graphs.

  The previous reduction also works for \textsc{Induced Disjoint $S$--$T$ Paths} as the invariant shown in \cref{clm:paths-to-sat} can be established in the same way under the weaker assumption that $G$ admits two mutually induced paths between $u_1^N$ and $\{u^N_{m+1}, u^S_{m+1}\}$, and between $u_1^S$ and $\{u^N_{m+1}, u^S_{m+1}\}$, respectively.
  In particular, the invariant shows that (if there is at least one clause) there cannot be $u_1^N$--$u^S_{m+1}$ and $u_1^S$--$u^N_{m+1}$ paths that are mutually induced in~$G$.  
  
  The reduction from \textsc{Clause-Linked Planar E3-OCC 3-SAT} of the proof of~\cref{thm:hardness-I2DP} is linear.
  Indeed it creates a~constant number of vertices for each variable and for each clause.
  By the Sparsification Lemma of Impagliazzo, Paturi, and Zane~\cite{sparsification} and classic linear reductions (see~\cite{Fellows95,Tippenhauer16}), $n$-variable \textsc{Clause-Linked Planar E3-OCC 3-SAT} requires time $2^{\Omega(\sqrt n)}$, under the ETH.
  We conclude that \textsc{Induced Disjoint $S$--$T$ Paths} with $|S|=|T|=2$ requires time $2^{\Omega(\sqrt n)}$ on $n$-variable string graphs of bounded twin-width and maximum degree.
\end{proof}

\section{Hardness of detecting a subcubic graph as an induced subdivision}\label{sec:isc-imc}

As a~consequence of the previous section, we can prove~\cref{thm:hardness-subcubic-subd}, which we recall.

\begin{reptheorem}{thm:hardness-subcubic-subd}
  \textsc{$H$-Induced Subdivision Containment} is NP-complete for the subcubic graph $H$ of~\cref{fig:H}.
\end{reptheorem}

\begin{proof}
  We reduce from \textsc{Induced 2-Disjoint Paths} in string graphs.
  Given any input $(G', s_1, t_1, s_2, t_2)$, we construct a~graph $G$ as follows.
  Take the disjoint union of $G'$ and~$H$, remove the edges $st$ and $s't'$ (of $H$), and identify $s_1$ with~$s$, $t_1$ with~$t$, $s_2$ with~$s'$, and $t_2$ with~$t'$.
  For the sake of clarity, for any $i \in [4]$, vertex set $A_i$ (in~$H$) is renamed $B_i$ in~$G$.
  Hence $V(G) = V(G') \cup \bigcup_{i \in [4]} B_i$.
  One can observe that if $(G', s_1, t_1, s_2, t_2)$ is a~positive instance of \textsc{Induced 2-Disjoint Paths}, then $H$ is an induced subdivision of $G$.
  We show that the converse also holds, hence this linear reduction is correct.

  Let $(\phi: V(H) \to V(G), (P_e)_{e \in E(H)})$ be an induced subdivision model of $H$ in $G$.
  We say that a~vertex of~$H$ is \emph{mapped} to its image by $\phi$.
  
  \begin{claim}\label{clm:anchor}
    For any $i \in [4]$, no vertex of $A_i$ can be mapped to a~vertex of $V(G')$.
  \end{claim}
  \begin{proofofclaim}
    No subdivision of $K_{3,3}$ has a~bridge.
    However, any induced subdivision in $G$ whose branching vertices intersect both $V(G')$ and $\bigcup_{i \in [4]} B_i$ contains a~bridge.
    This is because any of $s, t, s', t'$ is a~cutvertex in~$G$. 
    Thus for every $i \in [4]$, $\phi(A_i) \subset \bigcup_{i \in [4]} B_i$ or $\phi(A_i) \subset V(G')$.
    We now simply have to rule out the latter.
    Observe that there is no path in $G$ between two vertices in $V(G')$ that exits $V(G')$.
    Thus, if $\phi(A_i) \subset V(G')$, $G'$ would contain an induced subdivision of the 1-subdivision of $K_{3,3}$, which does not hold as $G'$ is a~string graph.
  \end{proofofclaim}

  \Cref{clm:anchor} implies that $\phi(\bigcup_{i \in [4]} A_i) = \bigcup_{i \in [4]} B_i$.
  Again, the absence of cutvertices in each $H[A_i]$ and $G[B_i]$ implies that for every $i \in [4]$ there is some $j \in [4]$ such that $\phi(A_i)=B_j$.
  As $|A_1| = |A_2| \neq |A_3| = |A_4|$ and $\phi$ is injective, $\{\phi(A_1), \phi(A_2)\}=\{B_1,B_2\}$ and $\{\phi(A_3), \phi(A_4)\}=\{B_3,B_4\}$.
  Now for the induced subdivision model of $H$ to be completed, there has to be in $G'$ two mutually induced paths between $s_1$ and $t_1$, and between $s_2$ and $t_2$.
  Thus $(G', s_1, t_1, s_2, t_2)$ is indeed a~positive instance.
\end{proof}

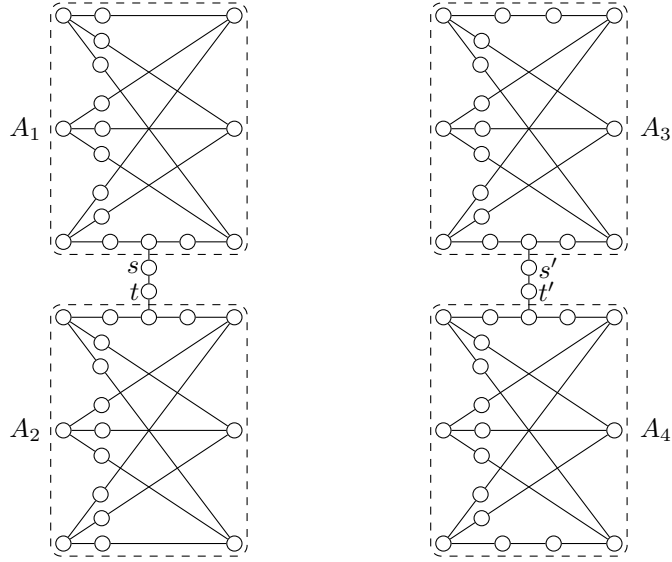
\begin{figure}[h!]
  \centering
  \begin{tikzpicture}[vertex/.style={draw,circle,inner sep=0.05cm, minimum width=0.2cm}]
    \def\s{1.5}

    \foreach \x/\y/\l in {0/0/1, 0/4/2}{
    \begin{scope}[xshift=\x cm, yshift = \y cm] 
      
    \foreach \i in {1,2,3}{
      \node[vertex] (u\l\i) at (0,\i * \s) {} ;
      \node[vertex] (v\l\i) at (1.5 * \s,\i * \s) {} ;
    }
    \node[draw,thin,dashed,rounded corners, inner sep=0.06cm, fit=(u\l1) (v\l3)] {} ;
    
    \foreach \i/\j in {1/2,1/3,2/1,2/2,2/3,3/1,3/2}{
      \path (u\l\i) to node[vertex, pos=0.2] (w\l\i\j) {} (v\l\j) ;
      \draw (u\l\i) -- (w\l\i\j) -- (v\l\j) ;
    }

    \end{scope}
    }

    \foreach \x/\y/\l in {5/0/3, 5/4/4}{
    \begin{scope}[xshift=\x cm, yshift = \y cm] 
      
    \foreach \i in {1,2,3}{
      \node[vertex] (u\l\i) at (0,\i * \s) {} ;
      \node[vertex] (v\l\i) at (1.5 * \s,\i * \s) {} ;
    }
    \node[draw,thin,dashed,rounded corners, inner sep=0.06cm, fit=(u\l1) (v\l3)] {} ;
    
    \foreach \i/\j in {1/2,1/3,2/1,2/2,2/3,3/1,3/2}{
      \path (u\l\i) to node[vertex, pos=0.2] (w\l\i\j) {} (v\l\j) ;
      \draw (u\l\i) -- (w\l\i\j) -- (v\l\j) ;
    }

    \end{scope}
    }
    % additional vertices
    \path (u43) to node[vertex, pos=0.33] (w433) {} (v43) ;
    \path (u43) to node[vertex, pos=0.66] (w433b) {} (v43) ;
    \draw (u43) -- (w433) -- (w433b) -- (v43) ;

    \path (u41) to node[vertex, pos=0.25] (w411l) {} (v41) ;
    \path (u41) to node[vertex, pos=0.75] (w411r) {} (v41) ;
    \path (u41) to node[vertex, pos=0.5] (w411) {} (v41) ;
    \draw (u41) -- (w411l) -- (w411) -- (w411r)  -- (v41) ;

    \path (u31) to node[vertex, pos=0.33] (w311) {} (v31) ;
    \path (u31) to node[vertex, pos=0.66] (w311b) {} (v31) ;
    \draw (u31) -- (w311) -- (w311b) -- (v31) ;

    \path (u33) to node[vertex, pos=0.25] (w333l) {} (v33) ;
    \path (u33) to node[vertex, pos=0.75] (w333r) {} (v33) ;
    \path (u33) to node[vertex, pos=0.5] (w333) {} (v33) ;
    \draw (u33) -- (w333l) -- (w333) -- (w333r) -- (v33) ;

    \path (u13) to node[vertex, pos=0.25] (w133l) {} (v13) ;
    \path (u13) to node[vertex, pos=0.5] (w133) {} (v13) ;
    \path (u13) to node[vertex, pos=0.75] (w133r) {} (v13) ;
    \draw (u13) -- (w133l) -- (w133) -- (w133r) -- (v13) ;
    
    \path (u21) to node[vertex, pos=0.25] (w211l) {} (v21) ;
    \path (u21) to node[vertex, pos=0.5] (w211) {} (v21) ;
    \path (u21) to node[vertex, pos=0.75] (w211r) {} (v21) ;
    \draw (u21) -- (w211l) -- (w211) -- (w211r) -- (v21) ;

    \path (u11) to node[vertex, pos=0.2] (w111) {} (v11) ;
    \draw (u11) -- (w111) -- (v11) ;

    \path (u23) to node[vertex, pos=0.2] (w233) {} (v23) ;
    \draw (u23) -- (w233) -- (v23) ;

    %\path[green!75!black] (v11) to node[vertex, pos=0.5] (mid) {} (u31) ;
    %\draw[green!75!black] (v11) -- (mid) -- (u31);

    % 2 edges
    \path (w133) to node[vertex, pos=0.3] (t) {} (w211) ;
    \path (w133) to node[vertex, pos=0.7] (s) {} (w211) ;
    \draw (w133) -- (t) -- (s)-- (w211) ;

    \path (w333) to node[vertex, pos=0.3] (tp) {} (w411) ;
    \path (w333) to node[vertex, pos=0.7] (sp) {} (w411) ;
    \draw (w333) -- (tp) -- (sp) -- (w411) ;

    \node[left] at (s) {$s$};
    \node[left] at (t) {$t$};
    \node[right] at (sp) {$s'$};
    \node[right] at (tp) {$t'$};
    %\node[above = 2pt] at (mid) {$m$};

    \node at (-0.5,7) {$A_1$} ;
    \node at (-0.5,2 * \s) {$A_2$} ;

    \node at (7.8,7) {$A_3$} ;
    \node at (7.8,2 * \s) {$A_4$} ;
    
  \end{tikzpicture}
  \caption{
  The graph $H'$ obtained from $H$ of~\cref{fig:H} by subdividing in each $A_i$ the two edges incident to the vertex with a~neighbor in $\{s,t,s',t'\}$.
  In total $H'$ has 74 vertices. 
  }
  \label{fig:Hp}
\end{figure}

The previous reduction also works for \textsc{$H$-Induced Minor Containment}.
However, the proof is slightly more involved.
We tune $H$ a~little bit such that it is still subcubic but has no two adjacent vertices of degree~3 (so that the forthcoming result answers a~question of Korhonen and Lokshtanov).
We call the resulting graph~$H'$; see~\cref{fig:Hp}.

\begin{reptheorem}{thm:hardness-subcubic-ind-minor}
  There is a~subcubic graph $H'$ such that every edge of $H'$ is incident to a~vertex of~degree~2, and \textsc{$H'$-Induced Minor Containment} is NP-complete, and requires time $2^{\Omega(\sqrt n)}$ on $n$-vertex graphs, unless the ETH fails.
\end{reptheorem}

\begin{proof}
  Again, we reduce from \textsc{Induced 2-Disjoint Paths} in string graphs and build from any input $(G', s_1, t_1, s_2, t_2)$, a~graph $G$ in the following way:
  Take the disjoint union of $G'$ and~$H'$, remove the edges $st$ and $s't'$ (of $H'$), and identify $s_1$ with~$s$, $t_1$ with~$t$, $s_2$ with~$s'$, and $t_2$ with~$t'$.
  For any $i \in [4]$, vertex set $A_i$ (in~$H'$) is renamed $B_i$ in~$G$.
  Thus $V(G) = V(G') \cup \bigcup_{i \in [4]} B_i$.
  If $(G', s_1, t_1, s_2, t_2)$ is a~positive instance of \textsc{Induced 2-Disjoint Paths}, then $H'$ is an induced subdivision, and hence an induced minor of~$G$.
  Again, we show that the converse also holds.

  Let $h := |V(H')| = 74$, and $(\mathcal M := \{X_1, \ldots, X_h\}, \phi: V(H') \to \mathcal M)$ be a~minimal induced minor model of $H'$ in~$G$.
First, assume that there is an~$i \in [4]$ and $x \in A_i$ such that $\phi(x) \cap V(G') \neq \emptyset$.
  As $G'$ is a~string graph (but $H'[A_i]$ is not), there is also some $y \in A_i$ and $j \in [4]$ such that $\phi(y) \cap B_j \neq \emptyset$.
  Let $u$ be the vertex of~$\{s_1, t_1, s_2, t_2\}$ with a~neighbor in $B_j$.
  As $u$ is a~cutvertex in $G$ that disconnects $B_j$ from the rest of~$G$, there is a~$z \in A_i$ such that $u \in \phi(z)$.

  As every branch set is connected, for every $z' \in A_i \setminus \{z\}$, either $\phi(z') \subset B_j$ or $\phi(z') \subset V(G-B_j)$.
  Furthermore, as $H'[A_i]$ has no cutvertex (in particular $z$ is not a~cutvertex of $H'[A_i]$), either for every $z' \in A_i \setminus \{z\}$, $\phi(z') \subset B_j$, or for every $z' \in A_i \setminus \{z\}$, $\phi(z') \subset V(G-B_j)$.
  The latter would contradict the minimality of~$(\mathcal M, \phi)$ as the (non-empty subset of) vertices of~$B_j$ could then be removed from every branch set; and in particular $\phi(y)$ would strictly decrease.

  Therefore, for every $z' \in A_i \setminus \{z\}$, $\phi(z') \subset B_j$.
  Moreover, it should hold that $z$ has a~neighbor in $\{s,t,s',t'\}$, say $s$.
  Now, one can change the induced minor model by moving $\phi(z) \cap V(G-B_j)$ to the adjacent branch set $\phi(s)$.
  This indeed still works as an induced minor model of $H'$ in $G$, and can greedily be made minimal (if need be).

  After at~most three more similar steps, we obtain a~(minimal) induced minor model $(\mathcal M', \phi')$ such that for every $x \in \bigcup_{i \in [4]} A_i$, $\phi'(x) \subseteq \bigcup_{i \in [4]} B_i$.
  We may then conclude as in the proof of~\cref{thm:hardness-subcubic-subd}.
  The ETH lower bound is a~direct consequence of~\cref{cor:tww} and of the fact that the present reduction is linear.
\end{proof}

\end{document}